\documentclass[12pt]{article}
\usepackage{epsfig}

\renewcommand{\thefootnote}{\fnsymbol{footnote}}

\begin{document}
\begin{flushright}
{\small hep-th/0405100}\\
{\small KAIST-TH 2004/04}
\end{flushright}
\vspace{0.6cm}

\begin{center}
{\Large \bf Fayet-Iliopoulos Terms in 5D Orbifold Supergravity}
\end{center}
\vspace{0.2cm}

\begin{center}
{\large
Hiroyuki~Abe\footnote{abe@hep.kaist.ac.kr}\,,~
Kiwoon~Choi\footnote{kchoi@hep.kaist.ac.kr}~ and
Ian-Woo~Kim\footnote{iwkim@hep.kaist.ac.kr} \\ \ \\
\normalsize \it Department of Physics,
Korea Advanced Institute of Science and Technology, \\
\normalsize \it Daejeon 305-701, Korea}
\end{center}
\vspace{0.2cm}

\begin{abstract}
\noindent
We discuss a locally supersymmetric formulation for the
boundary Fayet-Iliopoulos (FI) terms in 5-dimensional
$U(1)$ gauge theory on $S^1/Z_2$, using the four-form
multiplier mechanism to introduce the necessary $Z_2$-odd
FI coefficient. The physical consequence of the boundary
FI terms is studied within the full supergravity framework.
For both models giving a flat and a warped spacetime geometry,
the only meaningful deformation of vacuum configuration induced
by the FI terms is a kink-type vacuum expectation value of the
vector multiplet scalar field which generate a 5D kink mass for
charged hypermultiplet. This result for the four-form induced
boundary FI terms is consistent with the one
derived by the superfield formulation of 5D conformal supergravity.
\end{abstract}

\clearpage
\renewcommand{\thefootnote}{\arabic{footnote}}
\section{Introduction}
Theories with extra dimension can provide an attractive mechanism
to generate hierarchical structures in 4-dimensional (4D) physics
such as the weak to Planck scale hierarchy~\cite{Arkani-Hamed:1998rs,
Randall:1999ee} and the hierarchical Yukawa couplings of quarks and
leptons~\cite{Arkani-Hamed:1999dc}, as well as providing a new
mechanism to break grand unified symmetry~\cite{Kawamura:2000ev}
and/or supersymmetry~\cite{Scherk:1978ta,Antoniadis:1990ew,Randall:1998uk}.
In regard to generating the scale and/or Yukawa hierarchies,
quasi-localization of gravity~\cite{Randall:1999ee,Randall:1999vf}
and/or matter zero modes~\cite{Arkani-Hamed:1999dc} in extra dimension
is particularly interesting since it can generate exponentially
different 4D scales and/or Yukawa couplings even when the fundamental
parameters of the higher dimensional theory have similar magnitudes.
A simple theoretical framework to implement the idea of
quasi-localization would be 5D orbifold field theory on $S^1/Z_2$.
For instance, in 5D orbifold supergravity (SUGRA), the Randall-Sundrum
fine tuning~\cite{Randall:1999ee} of the bulk and brane cosmological
constants which is necessary for the gravity quasi-localization
can be naturally obtained by gauging the $U(1)_R$ symmetry with a
$Z_2$-odd gauge coupling~\cite{Choi:2002wx,Fujita:2001bd,Lin:2003ju}.
In the hypermultiplet compensator formulation of 5D off-shell
SUGRA~\cite{Fujita:2001bd}, this is equivalent to making the
compensator hypermultiplet to have a nonzero $Z_2$-odd gauge coupling
to the graviphoton. Also a nonzero 5D kink mass of matter hypermultiplet
causing the quasi-localization of matter zero mode can be obtained by
making the hypermultiplet to have a similar $Z_2$-odd gauge
coupling~\cite{Gherghetta:2000qt,Falkowski:2000er}.

It has been noted that globally supersymmetric 5D $U(1)$ gauge theory
allows  Fayet-Iliopoulos (FI) terms localized at the orbifold fixed
points~\cite{Ghilencea:2001bw,Barbieri:2002ic,GrootNibbelink:2002wv}.
In globally supersymmetric 4D theories, FI term is allowed for generic
$U(1)$, and  can lead to supersymmetry (SUSY) and/or gauge symmetry
breakings. However extending the 4D global SUSY to SUGRA severely
limits the possible FI terms. In 4D SUGRA, FI term is allowed only
when the associated $U(1)$ is either an $R$-symmetry~\cite{Barbieri:1982ac}
or a pseudo-anomalous $U(1)$ endowed with the
Green-Schwarz anomaly cancellation
mechanism~\cite{Green:sg}. On the other hand, in 5D orbifold SUGRA,
there  can be a boundary FI term~\cite{Barbieri:2002ic}
proportional to
$$\frac{1}{2}\partial_y\epsilon(y)=\delta(y)-\delta(y-\pi R)\,,$$
even when $U(1)$ is neither an $R$-symmetry nor a pseudo-anomalous
symmetry, where $\epsilon(y)$ is the periodic sign function on the
orbifold $S^1/Z_2$ whose fundamental domain is given by $0\leq y\leq \pi R$.
Such boundary FI terms generate a kink-type vacuum expectation
value (VEV) of the scalar component of $U(1)$ vector multiplet, thereby
giving a kink mass to $U(1)$-charged matter
hypermultiplets~\cite{Barbieri:2002ic,
GrootNibbelink:2002wv,Marti:2002ar,Abe:2002ps} causing the quasi-localization
of matter zero modes. It was also pointed out that such a boundary
FI term can be induced at one-loop level even when it is absent at tree
level~\cite{Ghilencea:2001bw,Barbieri:2002ic,
GrootNibbelink:2002wv,Marti:2002ar}.

In order to have a boundary FI term proportional to $\partial_y\epsilon(y)$
in 5D orbifold SUGRA,
one needs to introduce a $Z_2$-odd coupling $\xi_{FI}\epsilon(y)$.
It is in fact non-trivial to introduce a $Z_2$-odd coupling in 5D orbifold
SUGRA in a manner consistent with local supersymmetry. One known way is the
mechanism of Ref.~\cite{Bergshoeff:2000zn} in which the $Z_2$-odd factor
$\epsilon(y)$ appears as a consequence of the equations of motion of the
four-form multiplier field. This procedure does not interfere with local SUSY,
thus providing an elegant way to construct an orbifold SUGRA with $Z_2$-odd
couplings, starting from a theory only with $Z_2$-even couplings.

In this paper we wish to discuss a locally supersymmetric formulation for the
boundary FI terms in 5D $U(1)_X$ gauge theory on $S^1/Z_2$.
We apply the above mentioned four-form  mechanism to the known
formulation of 5D off-shell SUGRA~\cite{Fujita:2001bd}.
We consider a class of simple models with boundary FI terms, and analyze
the ground state solutions to examine whether the supersymmetry is broken or not
and also the generation of the hypermultiplet kink mass by the FI terms.
In Sec.~\ref{sec:2}, we derive the SUGRA action of bosonic fields containing
a boundary FI term proportional to $\partial_y\epsilon(y)$,
starting from the 5D off-shell SUGRA with four-form multiplier.

In Sec.~\ref{sec:3}, we derive the Killing spinor conditions and energy
functional in orbifold SUGRA models with FI terms for generic 4D
Poincare-invariant background geometry. We then examine in Sec.~\ref{sec:4}
the vacuum deformation caused by the boundary FI terms and the resulting
physical consequences.
We will find that for both models giving a flat and a warped spacetime
geometry, the vacuum solution preserves supersymmetry and
the only meaningful deformation of vacuum configuration induced
by the FI terms is a kink-type vacuum expectation value of the vector
multiplet scalar field which generate a 5D kink mass for charged
hypermultiplet. We then show that this result for the four-form induced
boundary FI terms is consistent with the one derived in \cite{Correia:2004pz} using the superfield
formulation of 5D conformal supergravity ~\cite{PaccettiCorreia:2004ri}
(see also \cite{Abe:2004ar}).

\section{5D orbifold supergravity with boundary FI terms}
\label{sec:2}

In this section, we construct the action of 5D orbifold SUGRA containing
the boundary FI terms for a bulk $U(1)$ gauge symmetry which is originally
neither
an $R$-symmetry nor a pseudo-anomalous symmetry. We will use the off-shell
formulation of 5D SUGRA on $S^1/Z_2$ which has been developed by Fujita,
Kugo and Ohashi~\cite{Fujita:2001bd}. In this formulation, gravitational
sector of the model is given by the  Weyl multiplet and the central charge
$U(1)_Z$ vector multiplet ${\cal V}_Z$ which contain the f\"unfbein $e_\mu^m$
and the $Z_2$-odd graviphoton $A^Z_\mu$, respectively, and a consistent
off-shell formulation is obtained by introducing a compensator hypermultiplet
${\cal H}_c$ \footnote{The theories with a single compensator hypermultiplet
can describe only the quaternionic hyperscalar manifolds $USp(2, 2n_H)/USp(2)
\times USp(2n_H)$ where $n_H$ is the number of physical hypermultiplets.
To describe other types of hyperscalar geometry, one needs to introduce
additional compensator hypermultiplets. However the physics of FI terms are
mostly independent of the detailed geometry of the hyperscalar manifold.
We thus limit the discussion to the theories with a single compensator
hypermultiplet.}. To discuss FI terms, we consider a minimal model with an
ordinary $U(1)_X$ vector multiplet ${\cal V}_X$ containing $Z_2$-even $U(1)$
gauge field $A^X_\mu$, and  also include a physical hypermultiplet ${\cal H}_p$.
As it is a theory on $S^1/Z_2$, all physical and nonphysical 5D fields have
the following boundary conditions:
$$
\Sigma(-y)=Z\Sigma(y)\,,
\quad \Sigma(y+2\pi R)=\Sigma(y)\,,
$$
where $Z^2=1$. It will be straightforward to extend our analysis to models
containing arbitrary number of vector multiplets and hypermultiplets with
more general boundary conditions.

In order  to introduce the $Z_2$-odd coupling $\xi_{FI}\epsilon(y)$ for
boundary FI terms, we apply the four-form  mechanism of
Ref.~\cite{Bergshoeff:2000zn}
to the 5D off-shell SUGRA~\cite{Fujita:2001bd}. For this purpose, we introduce
additional vector multiplet ${\cal V}_S$ as in Ref.~\cite{Fujita:2001bd}, so
our model contains three $U(1)$ vector multiplets at the starting point:
\begin{eqnarray}
\begin{array}{rcl}
{\cal V}_Z &=&
\big( \alpha,\, A^Z_\mu,\, \Omega^{Zi},\, Y^{Zij} \big)\,, \\
{\cal V}_X &=&
\big( \beta,\, A^X_\mu,\, \Omega^{Xi},\, Y^{Xij} \big)\,, \\
{\cal V}_S &=&
\big( \gamma, \, A^S_\mu, \,\Omega^{Si},\, Y^{Sij} \big)\,,
\end{array}
\nonumber
\end{eqnarray}
where $M^A=(\alpha,\beta,\gamma)$ ($A=Z,X,S$) are real scalar components,
$\Omega^i_A$ ($i=1,2$) are $SU(2)_U$-doublet symplectic Majorana spinors,
and $Y^{Aij}$ are $SU(2)_U$-triplet auxiliary components. Note that $\alpha$
is $Z_2$-even for $Z_2$-odd $A^Z_{\underline\mu}$, while $\beta$ and $\gamma$
are $Z_2$-odd for $Z_2$-even $A^X_{\underline\mu},A^S_{\underline\mu}$, and
the $Z_2$ transformation of $SU(2)_U$-doublet is given by $(\sigma_3)^i_j$.
Throughout this paper, $\mu=(\underline\mu,y)$ represents the 5D coordinate
directions with $\underline\mu$  representing the non-compact 4D coordinate
directions, and $m=0,1,2,3,4$ represents the 5D tangent space directions.

As for the hypermultiplets, we have
\begin{eqnarray}
\mbox{compensator}:\quad
{\cal H}_c &=&({\cal A}^x_i,\eta^x,{{\cal F}}^x_i)\,,
\nonumber \\
\mbox{physical}:\quad
{\cal H}_p &=& (\Phi^x_i,\zeta^x, F^x_i)\,,
\nonumber
\end{eqnarray}
where ${\cal A}^x_i,\Phi^x_i$ ($x=1,2$) are quaternionic hyperscalars,
$\eta^x,\zeta^x$ are symplectic Majorana hyperinos, and ${\cal F}^x_i,{F}^x_i$
are auxiliary components. In the following, we will use frequently a matrix
notation for hyperscalars, e.g.,
\begin{eqnarray}
\Phi &\equiv &
\left( \begin{array}{cc}
\Phi^{x=1}_{\ i=1} &
\Phi^{x=1}_{\ i=2} \\
\Phi^{x=2}_{\ i=1} &
\Phi^{x=2}_{\ i=2}
\end{array} \right) \ = \
\left( \begin{array}{cc}
\Phi_+ & \Phi_- \\
-\Phi_-^* & \Phi_+^*
\end{array} \right)\,,
\nonumber
\end{eqnarray}
where $\Phi_{\pm}$ are $Z_2$ parity eigenstates. In this matrix notation,
the symplectic reality condition and the $Z_2$ boundary condition are given by
\begin{eqnarray}
\Phi^*(y)=i\sigma_2\Phi(y)i\sigma_2^T\,,
\quad
\Phi(-y)=\sigma_3\Phi (y)\sigma_3\,.
\label{hyperboundary}
\end{eqnarray}

An off-shell formulation for the four-form mechanism to generate $Z_2$-odd
couplings in 5D orbifold SUGRA has been developed in
Ref.~\cite{Fujita:2001bd}.
In this formulation, the four-form multiplier $H_{\mu\nu\rho\sigma}$
corresponds to the dual of the auxiliary scalar component of a linear multiplet
${L}_H$. This dualization is defined under the background of the
Weyl multiplet
and a vector multiplet ${\cal V}_B$, and leads to a three-form field
$E_{\mu\nu\rho}$ as the dual of the constrained vector component of ${L}_H$.
Here we choose ${\cal V}_B$ to be the central charge vector multiplet
${\cal V}_Z$, and then the superconformal invariant couplings of ${L}_H$ and
${\cal V}_S$ include
\begin{eqnarray}
{\cal L}^{\rm bulk}_{\rm 4\textrm{-}form} &=&
e(Y^{Sij}-GY^{Zij})L_{ij}
\nonumber \\ &&\,
-\frac{1}{4!}\epsilon^{\lambda \mu \nu \rho \sigma} \left[ \left\{
F_{\lambda \mu}(A^S) -GF_{\lambda \mu}(A^Z) \right\} E_{\nu \rho \sigma}
+\frac{1}{2} G \partial_\lambda H_{\mu \nu \rho \sigma} \right]\,,
\label{eq:bkvpbrane}
\end{eqnarray}
where $G=M^S/M^Z=\gamma/\alpha$,
$L_{ij}$
is the $SU(2)_U$ triplet scalar component of the linear multiplet ${L}_H$,
$F_{\mu\nu}(A)$  is the field strength of the $U(1)$ gauge field $A^\mu$, and
$e=(-\mbox{det}(g_{\mu\nu}))^{1/2}$. Once we have the above bulk interactions,
$Z_2$-odd coupling constants can be obtained by introducing the following
superconformal invariant boundary Lagrangian density:
\begin{eqnarray}
{\cal L}^{\rm brane}_{\rm 4\textrm{-}form}=
\left( \Lambda_0 \delta(y)+\Lambda_\pi\delta(y-\pi R) \right)
\left[ \frac{1}{4!}
\epsilon^{\mu \nu \rho \sigma y} H_{\mu \nu \rho \sigma}
+e_{(4)}\alpha (i\sigma_3)^{ij}L_{ij} \right]\,,
\nonumber
\end{eqnarray}
where
$e_{(4)}=(-\mbox{det}(g_{\underline\mu\underline\nu}))^{1/2}$ for the induced
4D metric $g_{\underline\mu\underline\nu}$ on the boundaries. A detailed
derivation of the above 4-form Lagrangian densities can be found
in Ref.~\cite{Fujita:2001bd}.

In the above Lagrangian densities, $L_{ij},E_{\mu\nu\rho}$ and
$H_{\mu\nu\rho\sigma}$ play the role of Lagrangian multipliers.
By varying $H_{\mu \nu \rho \sigma}$, we obtain
\begin{eqnarray}
\partial_\mu G = -2\delta_\mu^y
\left( \Lambda_0 \delta(y)+\Lambda_\pi \delta(y-\pi R) \right)\,,
\nonumber
\end{eqnarray}
whose integrability condition leads to
\begin{eqnarray}
\Lambda_0 = -\Lambda_\pi\,.
\label{eq:integrability}
\end{eqnarray}
Taking the normalization $\Lambda_\pi =1$, one finds
\begin{eqnarray}
G=\gamma/\alpha = \epsilon(y)\,,
\label{eq:oddcoup}
\end{eqnarray}
where $\epsilon(y)=y/|y|$ is the periodic sign-function obeying
\begin{eqnarray}
&& \epsilon(y)=-\epsilon(-y)=\epsilon(y+2\pi R)=1
\quad (0<y<\pi R)\,,
\nonumber \\
&& \partial_y\epsilon(y)=2(\delta(y)-\delta(y-\pi R))\,.
\nonumber
\end{eqnarray}
Then the equations of motion for $E_{\mu \nu \rho}$ and $L_{ij}$ give
\begin{eqnarray}
&&F_{\mu \nu}(A^S) = \epsilon(y)F_{\mu \nu}(A^Z)\,,
\nonumber \\
&&Y^{Sij} = \epsilon(y)Y^{Zij}
+e^{-1}e_{(4)} \alpha (i\sigma_3)^{ij}(\delta(y)-\delta(y-\pi R))\,.
\label{eq:yrel}
\end{eqnarray}
Now using the relations (\ref{eq:oddcoup}) and (\ref{eq:yrel}), the redundant
vector multiplet ${\cal V}_S$ can be {\it replaced} by the central charge
vector multiplet ${\cal V}_Z$ multiplied by the $Z_2$-odd factor $\epsilon(y)$.
This four-form mechanism provides an elegant way to obtain a locally
supersymmetric theory of ${\cal V}_I$ ($I=Z,X$) involving $Z_2$-odd couplings,
starting from a locally supersymmetric theory of ${\cal V}_A$ ($A=Z,X,S$) and
the four-form multiplier multiplet involving only $Z_2$-even couplings.

The action of vector multiplets ${\cal V}_A$ ($A=Z,X,S$) is determined by the
norm function ${\cal N}$ which is a homogeneous cubic polynomial of the scalar
components $M^A=(\alpha,\beta,\gamma)$. As a minimal model incorporating the
boundary FI terms, we consider
\begin{eqnarray}
{\cal N}=C_{ABC}M^AM^BM^C=
\alpha^3 -\frac{1}{2} \alpha \beta^2
+\frac{1}{2} \xi_{FI} \alpha \beta \gamma\,.
\nonumber
\end{eqnarray}
Note that this form of ${\cal N}$ does not include any $Z_2$-odd coupling,
thus the results of Ref.~\cite{Fujita:2001bd} can be straightforwardly applied
for our ${\cal N}$. Then the auxiliary components $Y^{Aij}$ appear in the
bulk Lagrangian as
\begin{eqnarray}
e^{-1}{\cal L}_{Y} =
-\frac{1}{2}{\cal N}_{AB}Y^{Aij}Y^B_{ij}
+Y^A_{ij} {\cal Y}_A^{ij}\,,
\label{eq:laux}
\end{eqnarray}
where
\begin{eqnarray}
{\cal N}_{AB}&=&\frac{\partial^2 {\cal N}}{\partial M^A\partial M^B}\,,
\nonumber \\
{\cal Y}_A^{ij} &=& 2\left( {\cal A}^\dagger T_A {\cal A}
- \Phi^\dagger T_A \Phi \right)^{ij}\,,
\nonumber
\end{eqnarray}
for the $U(1)$ charge operators $T_A$.

In this paper, we limit the discussion to the case that $T_A$ commute with
the orbifolding $Z_2$ transformation $\Phi\rightarrow \sigma_3\Phi\sigma_3$.
Then under the condition that the model does not contain any $Z_2$-odd
coupling before the four-form multiplet is integrated out, the most general
form of the hyperscalar $U(1)$ charges consistent with the symplectic reality
condition (\ref{hyperboundary}) is given by
\begin{eqnarray}
\Big(\,T_Z,T_X,T_S\,\Big)\Phi&=&
\Big( 0,\,q,\,c \Big)i\sigma_3\Phi\,,
\nonumber \\
\Big(\,T_Z,T_X,T_S\,\Big){\cal A}&=&
\Big( 0,\,\tilde{q},\, \textstyle{-\frac{3}{2}}k \Big)i\sigma_3{\cal A}\,,
\label{eq:hypergauging}
\end{eqnarray}
where $q,c,\tilde{q},k$ are real constants.
If $U(1)_X$ is not an $R$-symmetry,
which is the case that we are focusing here,
the compensator hyperscalar ${\cal A}$
is neutral under $U(1)_X$, i.e.,
$$\tilde{q}=0\,.$$
When $\tilde{q}\neq 0$, so $U(1)_X$ is an $R$-symmetry,
there can be additional
FI terms both in the bulk and boundaries~\cite{nonintegrable:FI}.
As we will see,
in case that ${\cal A}$ has a nonzero $U(1)_S$
charge $\frac{3}{2}k$, the central
charge $U(1)_Z$ becomes an $R$-symmetry with a $Z_2$-odd gauge coupling
$\frac{3}{2}k\epsilon(y)$ after the four-form
multiplet is integrated out.
Such model has a ground state geometry being a slice of
${\rm AdS}_5$ with the AdS curvature $k$~\cite{Choi:2002wx,Fujita:2001bd},
so corresponds to the supersymmetric version of the Randall-Sundrum model.

Our goal is to derive the action of physical
fields by systematically integrating out
all non-physical degrees of freedom.
We already noted that the equations of motion of
$H_{\mu \nu \rho \sigma},E_{\mu \nu \rho}$ and $L_{ij}$ result
in the relations
(\ref{eq:oddcoup}) and (\ref{eq:yrel}).
Using these relations, we find first of all
\begin{eqnarray}
{\cal L}^{\rm bulk}_{\rm 4\textrm{-}form}
+{\cal L}^{\rm brane}_{\rm 4\textrm{-}form} =0 \,,
\nonumber
\end{eqnarray}
and
\begin{eqnarray}
e^{-1}{\cal L}_Y=
-\frac{1}{2}\tilde{\cal N}_{{I}{J}}
Y^{{I}ij}Y^{{J}}_{ij}
+Y^I_{ij}\tilde{\cal Y}^{ij}_{{I}}
+e^{-1}e_{(4)} \alpha (i\sigma_3)_{ij}{\cal Y}_S^{ij} (
\delta(y)-\delta(y-\pi R))\,,
\nonumber
\end{eqnarray}
where
\begin{eqnarray}
\tilde{\cal N}_{{I}{J}}&=&\frac{\partial^2\tilde{\cal N}}
{\partial M^I\partial M^J}\,,
\nonumber \\
\tilde{\cal Y}_I^{ij}&=&
2({\cal A}^\dagger {t}_I{\cal A}-\Phi^\dagger {t}_I\Phi)^{ij}
\nonumber \\ &&\,
-\frac{1}{2}\xi_{FI} e^{-1}e_{(4)}(i\sigma_3)^{ij}
(\alpha^2 \delta_I^X + \alpha \beta \delta_I^Z)
(\delta(y)-\delta(y-\pi R))\,,
\label{eq:s:onshellty} \\
{\cal Y}^{ij}_S&=&
-3ik({\cal A}^\dagger \sigma_3{\cal A})^{ij}-
2ic(\Phi^\dagger \sigma_3\Phi)^{ij}\,,
\nonumber
\end{eqnarray}
for $M^I=(\alpha,\beta)$ ($I=Z,X$) and
\begin{eqnarray}
\tilde{\cal N} =
\left.{\cal N}\right|_{\gamma=\epsilon(y)\alpha} \ = \
\tilde{C}_{IJK}M^IM^JM^K
=\alpha^3 -\frac{1}{2} \alpha \beta^2
+\frac{1}{2}\xi_{FI} \epsilon(y) \alpha^2 \beta\,.
\label{ntilde}
\end{eqnarray}
We remark that the $\xi_{FI}$-term in $\tilde{\cal N}$
is same as the one noted
in Ref.~\cite{Barbieri:2002ic}. Here the new $U(1)$ generators $t_I$ for
hyperscalars are given by
\begin{eqnarray}
\Big( {t}_Z,\,{t}_X \Big)\Phi &=&
\Big( c\epsilon(y),\, q \Big)i\sigma_3\Phi\,,
\nonumber \\
\Big( {t}_Z,\,{t}_X \Big){\cal A}&=&
\Big( \textstyle{-\frac{3}{2} k \epsilon(y)},\, 0 \Big)i\sigma_3{\cal A}\,,
\label{eq:u1charge}
\end{eqnarray}
and $\tilde{\cal N}^{{I}{J}}$ is the inverse matrix of $\tilde{\cal N}_{IJ}$.
Note that although ${\cal L}_Y$ was a bulk action in the original theory,
it gives boundary terms after the four-form multiplet is integrated out:
$$
e^{-1}{\cal L}_Y\,\,\rightarrow\,\,
{\rm tr}\big[i\sigma_3 (\xi_{FI}(\alpha^2Y^X+\alpha \beta Y^Z)/2
-\alpha{{\cal Y}}_S)\big]
e^{-1}e_{(4)}(\delta(y)-\delta(y-\pi R))\,.
$$

There appear additional boundary terms arising from the bulk kinetic term of
$M^A=(\alpha,\beta,\gamma)$ which is given by
\begin{eqnarray}
e^{-1}{\cal L}_{\rm kin}=-\frac{1}{4}{\cal N}
\frac{\partial^2\ln {\cal N}}{\partial M^A\partial M^B}
\nabla_m M^A \nabla^m M^B\,.
\nonumber
\end{eqnarray}
Using $\gamma=\epsilon(y)\alpha$, we find
\begin{eqnarray}
e^{-1}{\cal L}_{\rm kin} &=&
-\frac{1}{4}\,\tilde{\cal N}
\frac{\partial^2
\ln \tilde{\cal N}}{\partial M^I\partial M^J}
\nabla_{m}M^I\nabla^{m}M^J
\nonumber \\ &&
+\frac{1}{2}\,\xi_{FI}\alpha^2 \partial_4 \beta\,
e^{-1}e_{(4)} \Big(\delta(y)-\delta(y-\pi R)\Big)
+\Delta{\cal L}_{\rm brane}\,,
\label{braneterm}
\end{eqnarray}
where $M^I=(\alpha,\beta)$ and
\begin{eqnarray}
\Delta {\cal L}_{\rm brane}&=&
\frac{1}{4}\xi_{FI}
e^{-1}e_{(4)}\Big(\,
\delta(y)-\delta(y-\pi R)\,\Big)
\tilde{\cal N}^{-1}\Big[\,
(2\alpha^3\beta^2-\xi_{FI}\epsilon(y)\alpha^4\beta)\partial_4\beta
\nonumber \\ &&
-\,\, (4\alpha^4 \beta+\xi_{FI}\epsilon(y)\alpha^3 \beta^2) \partial_4\alpha
-\xi_{FI}\alpha^4\beta^2e^{-1}e_{(4)}(
\delta(y)-\delta(y-\pi R))\,\Big]\,.
\label{higherboundary}
\end{eqnarray}
If the $Z_2$-odd $\beta$ vanishes on the boundaries as was assumed
in~\cite{Fujita:2001bd,Bergshoeff:2000zn}, all operators in $\Delta {\cal L}_{\rm brane}$
would vanish also. However, as we will see, in the presence of the boundary FI terms,
$\beta$ develops a kink-type of fluctuation on the boundaries. In this situation, one
can {\it not} simply assume that operators involving $\beta$ vanish on the boundaries.
The values of $\Delta{\cal L}_{\rm brane}$ depend on how to regulate $\beta$ across
the boundary. This would cause a UV sensitive ambiguity in the theory, and makes it
difficult to find the correct on-shell SUSY transformation and Killing spinor conditions
on the boundaries.

It is in fact a generic phenomenon in orbifold field theory that boundary operators
can produce a UV sensitive singular behavior of bulk fields on the boundary.
Normally the resulting subtleties correspond to higher order effects in the perturbative
expansion in powers of dimensionful coupling constants which is equivalent to an expansion
in powers of $1/\Lambda$ for the cutoff scale $\Lambda$. Then the low energy physics below
$\Lambda$ can be described in a UV insensitive manner by limiting the analysis to an
appropriate order in the expansion. In our case, all the boundary terms appear in connection
with the dimensionful $Z_2$-odd coupling constants $\lambda=(\xi_{FI},k,c)$. For instance,
$\xi_{FI}\epsilon(y)$ leads to the integrable boundary FI terms, while the $Z_2$-odd gauge
couplings $\frac{3}{2}k\epsilon(y)$ and $c\epsilon(y)$ give rise to the integrable boundary
tensions and hyperscalar mass-squares. It turns out that at leading order in the perturbative
expansion in $\lambda=(\xi_{FI},k,c)$, those boundary terms generate the following
kink-type fluctuations
\begin{eqnarray}
\beta\,\sim\,\xi_{FI}\epsilon(y)\,,
\quad
\partial_yK\,\sim\, k\epsilon(y)\,,
\quad
\partial_y\Phi_+
\,\sim\, c\,\Phi_+(0)\epsilon(y)\,,
\label{kinkfluc}
\end{eqnarray}
for the spacetime metric
$g_{\underline\mu\underline\nu}=
e^{2K}\eta_{\underline\mu\underline\nu}$.
Then the precise value of boundary operators
involving
$\Omega_-=(\beta, \partial_y K,\partial_y\Phi_+)$
would depend on how to regulate the singular fixed point and the behavior of
$\Omega_-$ across the boundary. As in other cases, we can avoid this problem
of UV sensitivity by truncating the theory at an appropriate order in the
expansion in powers of $\lambda=(\xi_{FI},k, c)$.

To make this point more explicit, we parametrize $M^I=(\alpha,\beta)$ in terms
of a physical scalar field $\phi$ under the gauge fixing condition
$$\tilde{\cal N}=1$$
in the unit with the 5D Planck mass $M_5=1$:
\begin{eqnarray}
\alpha &=&
\frac{\cosh^{2/3} (\phi)}{(1+\xi_{FI}^2\epsilon^2(y)/8)^{1/3}}
\nonumber \\
&=&1+\frac{1}{3}\Big(
\phi^2-\frac{1}{8}\xi_{FI}^2\epsilon^2(y)\Big)
+{\cal O}(\phi^4)\,,
\nonumber \\
\beta &=&
\frac{\cosh^{2/3} (\phi)[ (2+\xi_{FI}^2\epsilon^2(y)/4)^{1/2}
\tanh (\phi) + \xi_{FI} \epsilon(y)/2 ]}{(
1+\xi_{FI}^2\epsilon^2(y)/8)^{1/3}}
\nonumber \\
&=& \frac{1}{2}\xi_{FI}\epsilon(y)+\sqrt{2}\phi+
{\cal O}(\phi^3)\,.
\nonumber
\end{eqnarray}
Then the very special manifold spanned by $\phi$ has the metric
\begin{eqnarray}
g_{\phi\phi}(\phi) &=& -\frac{1}{2} \frac{\partial^2\ln\tilde{\cal N}}
{\partial M^I\partial M^J}
\frac{\partial M^I}{\partial \phi}
\frac{\partial M^J}{\partial \phi}
\nonumber \\
&=&
\frac{1+2\cosh(2\phi)}{3\cosh^2(\phi)}
\,=\,1+\frac{1}{3}\phi^2+{\cal O}(\phi^4)\,.
\nonumber
\end{eqnarray}
Obviously, $\phi$ is $Z_2$-odd for $Z_2$-odd $\beta$. For $\xi_{FI}\neq 0$, $\phi$
has a kink-type fluctuation of ${\cal O}(\xi_{FI})$ near the boundaries.
Then, simply counting the powers of $\xi_{FI}$ in the limit $\xi_{FI}\ll 1$,
one easily finds $\alpha={\cal O}(1)$, $\beta={\cal O}(\xi_{FI})$,
$\partial_y\beta={\cal O}(\xi_{FI})$, and $\partial_y\alpha={\cal O}(\xi_{FI}^2)$
near the boundaries. With this power counting, the first boundary term in
(\ref{braneterm}) is of the order of $\xi_{FI}^2$, while
\begin{eqnarray}
\Delta {\cal L}_{\rm brane}={\cal O}(\xi_{FI}^4)\,.
\nonumber
\end{eqnarray}

In the next section, we will explicitly show that the energy functional
including only the boundary operators up to ${\cal O}(\xi_{FI}^2)$
takes the standard Bogomolny-squared form. This implies that the analysis
at ${\cal O}(\xi_{FI}^2)$ can be a consistent (approximate) framework to
examine the effects of the boundary FI terms in orbifold SUGRA.
If the operators of $\Delta {\cal L}_{\rm brane}$ are included, the
Killing spinor conditions will receive corrections of ${\cal O}(\xi_{FI}^3)$.
However, the values of $\Delta {\cal L}_{\rm brane}$ severely depend on the
way of regulating the singular functions $\epsilon(y)$ and $\delta(y)$
across the boundaries. It is thus expected that one needs informations on
the UV completion of orbifold SUGRA in order to make a complete analysis
including $\Delta {\cal L}_{\rm brane}$.

Upon ignoring the higher dimensional boundary terms of $\Delta{\cal L}_{\rm brane}$,
after integrating out all auxiliary fields {\it other than} $Y^I$ ($I=Z,X$),
we find the following Lagrangian density of bosonic fields:
\begin{eqnarray}
e^{-1}{\cal L}_{\rm bulk}&=&
-\frac{1}{2}R
-\frac{1}{4}\tilde{a}_{IJ}
F^I_{\mu \nu} F^{\mu \nu J}
+\frac{1}{2}\tilde{a}_{IJ}
\nabla^m M^I \nabla_m M^J
\nonumber \\
&&+
\frac{1}{8}e^{-1} \tilde{C}_{IJK}
\epsilon^{\lambda \mu \nu \rho \sigma} A^I_\lambda
 F^J_{\mu \nu} F^K_{\rho \sigma}
+{\rm tr} \Big[ |\nabla_m \Phi|^2-|\nabla_m {\cal A}|^2
-|V_{m}|^2
\nonumber \\ &&
-M^IM^J \big(
\Phi^\dagger {t}_I^\dagger {t}_J \Phi
-{\cal A}^\dagger {t}_I^\dagger {t}_J {\cal A}
\big) \Big]
\nonumber \\ &&
-\frac{1}{2} {\rm tr}\Big[
\tilde{\cal N}_{IJ} Y^{I\dagger}Y^J
-4Y^{I\dagger}
\left( {\cal A}^\dagger {t}_I {\cal A}
-\Phi^\dagger {t}_I \Phi \right)\Big] \,,
\nonumber \\
e_{(4)}^{-1} {\cal L}_{\rm brane} &=&  \Bigg[ \,
\frac{1}{2}\xi_{FI}\alpha^2
\Big(\, {\rm tr}[\,i\sigma_3 Y^X\,] +e^{-1}e_{(4)}\partial_y\beta \, \Big)
+\frac{1}{2}\xi_{FI}\alpha \beta\, {\rm tr}[\,i\sigma_3 Y^Z\,]
\nonumber \\ && \quad
-2\alpha\left(3k+\frac{3}{2}k\, {\rm tr} \left[ \Phi^\dagger \Phi \right]
+c\, {\rm tr} \left[ \Phi^\dagger \sigma_3 \Phi
\sigma_3 \right] \right) \Bigg]
\left( \delta(y)-\delta(y-\pi R) \right) \,.
\label{5daction}
\end{eqnarray}
where
\begin{eqnarray}
\tilde{a}_{{I}{J}} &=& -\frac{1}{2} \frac{\partial^2\ln\tilde{\cal N}}
{\partial M^I\partial M^J}\,,
\nonumber \\
V_{m} &=&
\frac{1}{2}\left( \Phi^\dagger (\nabla_m \Phi)
- (\nabla_m \Phi)^\dagger \Phi \right)\,.
\label{eq:s:onshelly}
\end{eqnarray}
Here the $2\times 2$ matrix valued compensator hyperscalar field can be chosen as
\begin{eqnarray}
{\cal A} &=&  \mathbf{1}_2\,
\sqrt{1+\frac{1}{2}{\rm tr} [\Phi^\dagger \Phi]}\,,
\label{eq:s:ufix}
\end{eqnarray}
which corresponds to one of the gauge fixing conditions in the hypermultiplet
compensator formulation of off-shell 5D SUGRA~\cite{Fujita:2001bd}.

The above action indeed includes the boundary FI term proportional to
$$
D_X=-\left( {\rm tr}[i\sigma_3Y^{X}]+e^{-1}e_{(4)}\partial_y\beta \right)\,,
$$
which can be identified as the $D$-component of the $N=1$ vector
superfield originating from the 5D vector multiplet ${\cal V}_X$.
According to the action (\ref{5daction}), the on-shell value of $Y^{I}$
is given by
\begin{eqnarray}
Y^{I}=\tilde{\cal N}^{IJ}\tilde{\cal Y}_J\,,
\label{onshell}
\end{eqnarray}
where $\tilde{\cal Y}_I$ are defined in (\ref{eq:s:onshellty}).
Then after integrating out the auxiliary components $Y^I$, the full 5D scalar
potential is given by
\begin{eqnarray}
V_{5D}&=& \,{\rm tr}\Big[
M^IM^J \big\{ \Phi^\dagger {t}_I^\dagger {t}_J \Phi
-{\cal A}^\dagger {t}_I^\dagger {t}_J {\cal A} \big\}
-\frac{1}{2} \tilde{\cal N}^{IJ}
\tilde{\cal Y}_I^\dagger \tilde{\cal Y}_J \Big]
\nonumber \\
&&+\, 2e^{-1}e_{(4)} \alpha \left(
3k+\frac{3}{2} k\,{\rm tr} \left[ \Phi^\dagger \Phi \right]
+c\,{\rm tr} \left[ \Phi^\dagger \sigma_3 \Phi
\sigma_3\right] \right) \left(
\delta (y)-\delta(y-\pi R)\right)\,.
\label{eq:s:sp0}
\end{eqnarray}

With the above results, one easily finds
\begin{eqnarray}
V_{5D}=-6k^2+6ke^{-1}e_{(4)}(\delta(y)-\delta(y-\pi R))+\ldots\,,
\nonumber
\end{eqnarray}
where the ellipsis stands for field-dependent terms. This shows that a nonzero $U(1)_Z$
charge of the compensator hyperscalar, i.e., $\frac{3}{2}k\epsilon(y)\sigma_3$, gives rise
to the correctly tuned bulk and boundary cosmological constants yielding a warped
Randall-Sundrum geometry with AdS curvature $k$. After the compensator gauge fixing
(\ref{eq:s:ufix}), this $U(1)_Z$ charge corresponds to a $U(1)_R$ gauge charge.
As a result, the graviphoton $A^Z_\mu$ becomes a $U(1)_R$ gauge field and its auxiliary
component $Y^Z$ has a bulk FI term $\sim k\epsilon(y)Y^Z$ which leads to the negative bulk
cosmological constant in $V_{5D}$.

Also from the on-shell expression of the $N=1$ auxiliary components, e.g., the order parameter
$F$ in Eq.~(\ref{killingcondition}), one can see that the hypermultiplet obtains the following
5D mass determined by the $U(1)_I$ ($I=Z,X$) charges and the $U(1)_X$ FI coefficient:
\begin{eqnarray}
m_{\Phi_\pm}^2 &=&
\Big( c_w^2 \pm c_w - {\textstyle \frac{15}{4}} \Big) k^2 \epsilon^2(y)
+(3 \mp 2c_w)k \big( \delta(y)-\delta(y-\pi R) \big)\,,
\nonumber \\
m_{\zeta} &=& c_wk \epsilon(y)\,,
\label{eq:hmass}
\end{eqnarray}
where we consider only the hyperscalar mass from the $F$-term potential,
$\lambda = (\xi_{FI},k,c)$, and
$$
c_wk \epsilon(y) \equiv -\Big( c\epsilon(y)
+q \langle \beta(\phi) \rangle \Big).
$$
Note that this hypermultiplet mass reproduces  the result of
Ref.~\cite{Gherghetta:2000qt}.

\section{Killing spinor conditions and energy functional}
\label{sec:3}

In this section, we derive the Killing spinor equations and the energy functional
in 5D orbifold SUGRA model with boundary FI terms for generic 4D Poincare invariant
metric configuration:
\begin{eqnarray}
ds^2=e^{2K(y)} \eta_{\underline\mu \underline\nu} (x)
dx^{\underline\mu} dx^{\underline\nu}-dy^2\,.
\label{eq:s:bkg}
\end{eqnarray}
We will employ the relation $e^y_4=e^{-1}e_{(4)}=1$ frequently in the following.
Applying the local SUSY transformations of the gravitino $\psi^i_\mu$, the gauginos
$\Omega^i_I$, and the compensator and physical hyperinos
$\eta^x,\zeta^x$, we find~\cite{Fujita:2001bd}
\begin{eqnarray}
\delta \psi_\mu^i &=&
\left( \partial_\mu
-\frac{1}{4} \omega_\mu^{ab} \gamma_{ab} \right) \varepsilon^i
-\frac{1}{2}\delta^y_\mu(\Phi^\dagger\partial_y\Phi
-\partial_y\Phi^{\dagger}\Phi)^i_{\ j} \varepsilon^j
-\frac{1}{3}\frac{\tilde{\cal N}_{{I}}}{\tilde{\cal N}}
(Y^{{I}})^i_{\ j} \gamma_\mu \varepsilon^j\,,
\nonumber \\
\delta \Omega^{i{I}} &=&
-\frac{i}{2} \gamma_5 \partial_y M^{{I}} \varepsilon^i
+(Y^{{I}})^i_{\ j} \varepsilon^j
-\frac{1}{3}M^{{I}}
\frac{\tilde{\cal N}_{{J}}}{\tilde{\cal N}}
(Y^{{J}})^i_{\ j} \varepsilon^j,
\nonumber\\
\delta\eta^x&=&\left( i\gamma_5
\partial_y-\frac{1}{2}i\gamma_5(\Phi^\dagger\partial_y\Phi
-\partial_y\Phi^\dagger\Phi)_{ij}{\cal A}^{xj}
-M^I (t_I)^x_{x'}
{\cal A}^{x'}_{\ i} \right) \varepsilon^i
+{\cal A}^x_{\ i}
\frac{\tilde{\cal N}_{{I}}}{\tilde{\cal N}}
(Y^{{I}})^i_{\ j} \varepsilon^j\,,
\nonumber \\
\delta\zeta^x&=&\left( i\gamma_5
\partial_y-\frac{1}{2}i\gamma_5(\Phi^\dagger\partial_y\Phi
-\partial_y\Phi^\dagger\Phi)_{ij}{\Phi}^{xj}
-M^I (t_I)^x_{x'}
{\Phi}^{x'}_{\ i} \right) \varepsilon^i
+{\Phi}^x_{\ i}
\frac{\tilde{\cal N}_{{I}}}{\tilde{\cal N}}
(Y^{{I}})^i_{\ j} \varepsilon^j\,,
\label{eq:app:hpsusytrs}
\end{eqnarray}
where $\tilde{\cal N}_I=\partial \tilde{\cal N}/\partial M^I$,
${\cal A}^x_i$ and $Y^I$ are given by (\ref{eq:s:ufix}) and (\ref{onshell}),
respectively, and we use the convention $i\gamma_5=\gamma^{\mu=y}$. Here the
local SUSY transformation spinor $\varepsilon^i$ obeys the $Z_2$-orbifolding
condition $\varepsilon^i(-y) = \gamma_5 (\sigma_3)^i_{\ j} \varepsilon^j(y)$.
From the above local SUSY transformations, one can find that the Killing spinor
conditions for 4D Poincare invariant spacetime are given by
\begin{eqnarray}
&&\Big(\partial_y -
\frac{1}{2} \left( \Phi^{\dagger} \partial_y{\Phi}
-\partial_y{\Phi}^\dagger \Phi \right)
+\frac{1}{6}i M^{{I}} \tilde{\cal Y}_{{I}}
\gamma_5\Big)\, \varepsilon = 0\,,
\nonumber \\
&&\Big(\partial_y{K}
+ \frac{1}{3}i M^{{I}} \tilde{\cal Y}_{{I}}
\gamma_5 \Big)\,\varepsilon = 0\,,
\nonumber \\
&&
\Big(-\frac{1}{2} \partial_y{M}^{{I}}
-i \Big( \tilde{\cal N}^{IJ}
-\frac{1}{6}M^IM^J \Big)
\tilde{\cal Y}_{{J}}
\gamma_5\Big) \,\varepsilon = 0\,,
\nonumber \\
&&\Big( \partial_y{\Phi}- \frac{1}{2}
\Phi \big( \Phi^{\dagger} \partial_y{\Phi}
-\partial_y{\Phi}^\dagger \Phi \big)^\dagger
 +i M^I t_I\Phi \gamma_5
-\frac{1}{2}i M^{{I}} \Phi \tilde{\cal Y}_{{I}}
\gamma_5 \Big)\,\varepsilon = 0\,,
\nonumber \\
&&\Big( {\partial_y{\cal A}}- \frac{1}{2}
{\cal A} \big( \Phi^{\dagger} \partial_y{\Phi}
-\partial_y{\Phi}^\dagger \Phi \big)^\dagger
+i M^It_I{\cal A} \gamma_5
-\frac{1}{2}i M^{{I}}{\cal A} \tilde{\cal Y}_{{I}}
\gamma_5\Big) \,\varepsilon = 0\,.
\label{eq:app:cpke}
\end{eqnarray}

These Killing spinor conditions can be rewritten as
\begin{eqnarray}
&&\partial_y{\varepsilon}_+
+\left( \frac{i}{6} M^{{I}} \tilde{\cal Y}_{{I}}\sigma_3
-\frac{1}{2} \big( \Phi^{\dagger} \partial_y \Phi
-\partial_y \Phi^\dagger \Phi \big) \right)
\varepsilon_+ \ = \ 0\,,
\nonumber \\
&& \kappa \ \equiv \ \partial_y{K} +\frac{i}{3} M^{{I}}
\tilde{\cal Y}_{{I}} \sigma_3 \ = \ 0\,,
\nonumber \\
&& G^I \ \equiv \
\tilde{\cal N}^{IJ} \Big(
\partial_y \tilde{\cal N}_J
-3(\partial_y \tilde{C}_{JKL})M^KM^L
+2i\tilde{\cal Y}_{{J}} \sigma_3
-\frac{2i}{3}\tilde{\cal N}_JM^K \tilde{\cal Y}_{{K}}
\sigma_3 \Big) \ = \ 0\,,
\nonumber \\
&& F \ \equiv \ \partial_y{\Phi} - \frac{1}{2}
\Phi \big( \Phi^{\dagger} \partial_y{\Phi}
-\partial_y{\Phi}^\dagger \Phi \big)^\dagger
+iM^I t_I \Phi \sigma_3
-\frac{i}{2}\Phi M^{{I}} \tilde{\cal Y}_{{I}}\sigma_3 \ = \ 0\,,
\nonumber \\
&& {\cal F} \ \equiv \ \partial_y{{\cal A}}-\frac{1}{2}
{\cal A} \big( \Phi^{\dagger} \partial_y{\Phi}
-\partial_y{\Phi}^\dagger \Phi \big)^\dagger
+iM^I t_I {\cal A} \sigma_3
-\frac{i}{2}{\cal A}M^{{I}}\tilde{\cal Y}_{{I}}\sigma_3 \ = \ 0\,,
\label{killingcondition}
\end{eqnarray}
where $\tilde{\cal Y}_I$ are given by (\ref{eq:s:onshellty}),
$\varepsilon=(\varepsilon^{i=1},\varepsilon^{i=2})$, and
$\varepsilon_+=\frac{1}{2}(1+\gamma_5\sigma_3)\varepsilon$.
In the above, $\kappa=0$ corresponds to the gravitino Killing condition, $G^I=0$ is
the gaugino Killing condition, $F=0$ is the physical hyperino Killing condition,
and ${\cal F}=0$ comes from the compensator hyperino Killing condition. In fact,
${\cal F}=0$ is not independent from other Killing conditions since the compensator
hypermultiplet is not a physical degree of freedom. However, here we treat it
separately for later convenience.
Here we remark that $G^I=0$ ($I=Z,X$) provide {\it two}  independent Killing conditions
even though there is only one physical gaugino.
This is essentially due to the existence of the second term in the
relation
$\partial_y M^I=
\frac{\partial M^I}{\partial \phi} \partial_y \phi
+\frac{\partial M^I}{\partial \epsilon(y)} \partial_y \epsilon(y)$
after the gauge fixing $\tilde{\cal N}=1$ \cite{Correia:2004pz}, which
results from the fact that the norm function $\tilde{\cal N}$
carries the $y$-dependent coefficient $\epsilon(y)$ induced
by the four-form mechanism.
In the presence of this extra term, even though we have
the relation $\tilde{\cal N}^{IJ}-\frac{1}{6}M^IM^J
=-\frac{1}{2}g^{\phi \phi}
\frac{\partial M^I}{\partial \phi}
\frac{\partial M^J}{\partial \phi}$ for the  gaugino Killing equation of (\ref{eq:app:cpke}),
we can not factorize out the factor $\frac{\partial M^I}{\partial\phi}$
from the gaugino Killing equation, thus there are two independent
gaugino Killing conditions.

When the above Killing conditions are all satisfied, the Killing spinor is given by
\begin{eqnarray}
\varepsilon_+(y) &=&
\exp\left[\frac{1}{2}(K(y)-K(0))
+\frac{1}{2}\int_0^y dz\,(\Phi^\dagger\partial_z\Phi
-\partial_z\Phi^{\dagger}\Phi)\right] \varepsilon_+(0)\,.
\nonumber
\end{eqnarray}
The boundary FI terms affect the gaugino Killing condition $G^I=0$
through $\tilde{\cal Y}_X\partial\beta/\partial\phi$, thus gives
an effect of ${\cal O}(\xi_{FI})$ to the Killing condition.
On the other hand, the other Killing conditions
are affected through $\beta\tilde{\cal Y}_X$ which
gives an effect of ${\cal O}(\xi_{FI}^2)$ for $\beta
={\cal O}(\xi_{FI})$.

In fact, the above Killing conditions are not sufficient conditions for $N=1$ SUSY
preserving 4D Poincare invariant vacuum. The solution should be a stationary point
of the energy functional with a vanishing vacuum energy density.
For the 4D
Poincare invariant geometry (\ref{eq:s:bkg}),
the energy functional resulting from the 5D action (\ref{5daction})
is given by
\begin{eqnarray}
E &=& \int dy \,\, {\cal E}\,,
\nonumber
\end{eqnarray}
where
\begin{eqnarray}
{\cal E}&=&e^{4K}\Bigg[
4\partial_y^2K+10(\partial_yK)^2
+\frac{1}{2}\tilde{a}_{IJ}
(\partial_y M^I)(\partial_y M^J)
+{\rm tr}\Bigg\{
\partial_y\Phi^\dagger \partial_y \Phi
-\partial_y{\cal A}^\dagger\partial_y{\cal A}
\nonumber \\
&&\quad\quad+\,\frac{1}{4}|\partial_y\Phi^\dagger
\Phi-\Phi^\dagger\partial_y\Phi|^2
+M^IM^J \big( \Phi^\dagger t_I^\dagger t_J \Phi
-{\cal A}^\dagger t_I^\dagger t_J {\cal A} \big)
-\,\frac{1}{2} \tilde{\cal N}^{IJ}
\tilde{\cal Y}_I^\dagger \tilde{\cal Y}_J
\nonumber \\
&&\quad\quad +\Big\{ 2\alpha \sigma_3
\Big( \frac{3}{2}k\,{\cal A}^\dagger\sigma_3{\cal A}
+c\,\Phi^\dagger\sigma_3\Phi \Big)
-\frac{1}{4}\xi_{FI} \alpha^2 \partial_y\beta\,\mathbf{1}_2 \Big\}
(\delta(y)-\delta(y-\pi R))\,\Bigg\}  \Bigg]\,,
\label{eq:sp0}
\end{eqnarray}
for ${\cal A}=(1+\frac{1}{2}{\rm tr}(\Phi^\dagger\Phi))^{1/2}
\mathbf{1}_2$.
It is then straightforward to find that
\begin{eqnarray}
{\cal E}
&=& e^{4K}\,\Bigg[\,
\frac{1}{2}\tilde{a}_{IJ}G^{I \dagger} G^J-6|\kappa|^2
+{\rm tr}\Bigg\{|F|^2-|{\cal F}|^2
-\frac{1}{2}|\partial_y\Phi^\dagger
\Phi-\Phi^\dagger\partial_y\Phi|^2
\nonumber \\
&&\quad\quad
-\frac{1}{2}iM^I\left(\Phi^\dagger\partial_y\Phi
-\partial_y\Phi^\dagger\Phi \right)\left[
{\cal A}^\dagger t_I
{\cal A}-\Phi^\dagger t_I\Phi,\, \sigma_3 \right]\Bigg\}
\Bigg]
\nonumber \\
&&+\,\partial_y\Bigg[\,
\frac{i}{2}e^{4K}M^I
\,{\rm tr}\,
\sigma_3\Big({\cal A}^\dagger t_I{\cal A}-\Phi^\dagger t_I\Phi
\Big)\, +\textrm{h.c.}
+4e^{4K}\partial_yK \,\Bigg]\,,
\nonumber
\end{eqnarray}
so the energy functional can be rewritten as
\begin{eqnarray}
E &=& \int \,dy\, e^{4K} \Bigg(\,
\frac{1}{2}\tilde{a}_{IJ}G^{I \dagger}G^J-6|\kappa|^2+
{\rm tr}\Big\{\,|F|^2-|{\cal F}|^2
-\frac{1}{2}|\Phi^\dagger\partial_y\Phi
-\partial_y\Phi^\dagger\Phi|^2\Big\}
\nonumber \\
&&\quad\quad\quad\quad\quad\,
-\frac{1}{2}iM^I{\rm tr}\Big\{ \,(\Phi^\dagger\partial_y\Phi
-\partial_y\Phi^\dagger\Phi)
\left[ {\cal A}^\dagger t_I{\cal A}
-\Phi^\dagger t_I\Phi,\, \sigma_3 \right]\,
\Big\}\,\Bigg)\,.
\nonumber
\end{eqnarray}
To arrive at this Bogomolny form starting from the 5D action
(\ref{5daction}), we have truncated  the boundary operators of
${\cal O}(\lambda\xi_{FI}^2)$ ($\lambda=(\xi_{FI},k,c)$) which
are UV sensitive as their values depend on the way of regulating
the kink fluctuations (\ref{kinkfluc}).

The hyperscalar part of the above energy functional is not written in the
more familiar on-shell form~\cite{Ceresole:2001wi}. We confirmed that our
energy functional can be rewritten in the standard on-shell form involving
the metric of the quaternionic hyperscalar manifold $USp(2,2)/USp(2)\times USp(2)$.
It is in fact more convenient to use the above form of the hyperscalar energy
functional rather than the conventional on-shell form to discuss supersymmetric
solutions. For instance, it immediately shows that a  field configuration
satisfying  the Killing conditions (\ref{killingcondition}) and also the
stationary conditions,
\begin{eqnarray}
&&\Phi^\dagger\partial_y\Phi-\partial_y\Phi^\dagger\Phi=0\,,
\nonumber \\
&&[\,\sigma_3, {\cal A}^\dagger t_I{\cal A}-\Phi^\dagger t_I
\Phi\,]=0\,,
\label{stationarycondition}
\end{eqnarray}
corresponds to a supersymmetric solution with vanishing vacuum energy.
A simple solution of the above stationary conditions  is
\begin{eqnarray}
\Phi =
\left( \begin{array}{cc}
v(y) & 0 \\
0 & v(y)
\end{array} \right) \,,
\label{hypervev}
\end{eqnarray}
where $v$ is a real VEV. For this form of $\Phi$, one easily finds
\begin{eqnarray}
F=\mathbf{1}_2\,f(y)\,,\quad
{\cal F}=\mathbf{1}_2\,\frac{v(y)}{\sqrt{1+v^2(y)}}
\big( f(y)+v^{-1}(y)\Delta(y) \big)\,,
\nonumber
\end{eqnarray}
where
\begin{eqnarray}
f(y)&=&\partial_y v-
v(1+v^2)\Big[
\,\Big(\frac{3}{2}k+c\Big)\epsilon(y)\alpha(\phi)
+q\beta(\phi)\,\Big] +v\Delta(y)\,,
\nonumber \\
\Delta(y) &=&
-\frac{1}{2}\xi_{FI}\alpha^2(\phi)\beta(\phi) \,
\Big(\delta(y)-\delta(y-\pi R)\Big)\,,
\nonumber
\end{eqnarray}
for which the energy functional is given by
\begin{eqnarray}
E=\int \,dy \,\, e^{4K}\Big(\,
\frac{1}{2} \tilde{a}_{IJ}G^{I \dagger}G^J
-6|\kappa|^2+\frac{2}{1+v^2}|f|^2\,\Big)\,,
\label{5denergy}
\end{eqnarray}
where again we have truncated the UV sensitive boundary operators
of ${\cal O}(\lambda \xi_{FI}^2)$ ($\lambda=(\xi_{FI},k,c)$).

\section{Vacuum deformation induced by boundary FI terms}
\label{sec:4}

In this section, we study the deformation of vacuum solution
induced by the boundary FI terms. We will see that
the only meaningful vacuum deformation induced by the boundary FI terms
in 5D orbifold SUGRA is a kink-type VEV of the vector multiplet scalar:
$\langle \beta \rangle={\cal O}(\xi_{FI})$, which
is essentially same as the vacuum deformation in rigid
SUSY limit.

To proceed, we assume that the hyperscalar VEV is
given by (\ref{hypervev}) for which the stationary conditions (\ref{stationarycondition})
are automatically satisfied. Then from
\begin{eqnarray}
\tilde{\cal Y}_Z&=&2i\epsilon(y)\left[\,-\frac{3}{2}k
\Big(1 + \frac{1}{2}{\rm tr}(\Phi^\dagger \Phi) \Big)\sigma_3
-c\,\Phi^\dagger \sigma_3 \Phi\,\right]
-\frac{i}{2}\xi_{FI}\alpha \beta \sigma_3
(\delta(y)-\delta(y-\pi R))\,,
\nonumber \\
\tilde{\cal Y}_X&=&
-2iq\Phi^\dagger\sigma_3\Phi
-\frac{i}{2}\xi_{FI}\alpha^2\sigma_3
(\delta(y)-\delta(y-\pi R))\,.
\nonumber
\end{eqnarray}
one easily finds that the Killing conditions
(\ref{killingcondition}) give
\begin{eqnarray}
\varepsilon_+(y)&=&
e^{(K(y)-K(0))/2}\varepsilon_+(0)\,,
\nonumber \\
\partial_yK&=& -k\epsilon(y)\alpha(\phi)
-\frac{2}{3}v^2\left[
\left( \frac{3}{2}k+c \right)\epsilon(y)\alpha(\phi)
+q\beta(\phi)\right]+\frac{2}{3}\Delta(y)\,,
\nonumber \\
\partial_y \tilde{\cal N}_I &=&
-2\tilde{\cal N}_I
\bigg\{-k\epsilon(y)\alpha(\phi)
-\frac{2}{3}v^2 \left[ \left( \frac{3}{2}k+c \right)
\epsilon(y) \alpha(\phi)
+q\beta(\phi) \right] +\frac{2}{3}\Delta(y) \bigg\}
\nonumber \\ &&
-6k\epsilon(y)\delta_I^{\ Z}
-4\left\{ \left( \frac{3}{2}k+c \right) \epsilon(y)\delta_I^{\ Z}
+q\delta_I^{\ X} \right\} v^2
+2\alpha^{-1}(\phi) \Delta(y) \delta_I^{\ Z}\,.
\nonumber \\
\partial_yv&=& v(1+v^2)\left[\,
\left(\frac{3}{2}k+c \right)\epsilon(y)
\alpha(\phi)+q\beta(\phi) \,\right] -v\Delta(y)\,.
\label{eq:keqsads}
\end{eqnarray}
Combining the second and third equations of
(\ref{eq:keqsads})  together, we also obtain
\begin{eqnarray}
e^{-2K} \partial_y (e^{2K} \tilde{\cal N}_I)=
-6k\epsilon(y)\delta_I^{\ Z}
-4\left\{ \left( \frac{3}{2}k+c \right) \epsilon(y)\delta_I^{\ Z}
+q\delta_I^{\ X} \right\} v^2
+2\alpha^{-1}(\phi) \Delta(y) \delta_I^{\ Z}\,.
\label{eq:gakcads}
\end{eqnarray}

Obviously, the last equation of (\ref{eq:keqsads})
is satisfied by
$$
v=0\,.
$$
For such hyperscalar vacuum value,
Eq.~(\ref{eq:gakcads}) for $I=X$ leads to
$\tilde{\cal N}_X=-\alpha \beta+\xi_{FI}\epsilon(y) \alpha^2/2=0$, which results in
\begin{eqnarray}
\beta(\phi) &=& \frac{1}{2}\xi_{FI}\epsilon(y)\alpha(\phi)\,.
\label{eq:betavev}
\end{eqnarray}
Then $\tilde{\cal N}=1$ gives
$\alpha(\phi)=(1+\xi^2_{FI} \epsilon^2(y)/8)^{-1/3}$, and thus
$$
\langle \phi \rangle = 0\,.
$$
With these VEVs of $\phi$ and $\Phi$, the $I=Z$ component of Eq.~(\ref{eq:gakcads}) is given by
\begin{eqnarray}
2\partial_y K
&=& -6k\epsilon(y) \tilde{\cal N}_Z^{-1}
-\partial_y \ln \tilde{\cal N}_Z
-\xi_{FI} \tilde{\cal N}_Z^{-1} \alpha(\phi) \beta(\phi)
\big( \delta(y)-\delta(y-\pi R) \big)
\nonumber \\ &=&
-2k\epsilon(y) \Big( 1+ \frac{1}{8}\xi^2_{FI} \epsilon^2(y) \Big)^{-1/3}
-\frac{1}{3} \Big( 1+ \frac{1}{8}\xi^2_{FI} \epsilon^2(y) \Big)^{-1}
\xi_{FI}^2 \epsilon(y) \big( \delta(y)-\delta(y-\pi R) \big)\,.
\nonumber
\end{eqnarray}
Upon ignoring the small fluctuation of $K$
on the boundary of order $\xi_{FI}^2$, the above equation gives
$$
K \simeq -k|y|\,,
$$
which yields the AdS$_5$ geometry:
\begin{eqnarray}
ds^2=e^{-2k|y|}\eta_{\underline\mu\underline\nu}
dx^{\underline\mu}dx^{\underline\nu}-dy^2\,,
\label{adsmetric}
\end{eqnarray}
and the corresponding Killing spinor:
$$
\varepsilon_+(y)=e^{-k|y|/2}\varepsilon_+(0)\,.
$$

In the above, we found that at the order  of $\xi_{FI}$
the only vacuum deformation
induced by the FI terms  is
\begin{eqnarray}
\langle \beta \rangle
&=& \frac{1}{2}\xi_{FI}\epsilon(y) \alpha(\phi=0)
\ = \ \frac{1}{2}\xi_{FI}\epsilon(y) +{\cal O}(\xi_{FI}^2)\,,
\nonumber
\end{eqnarray}
irrespective of the value of $k$.
One interesting consequence of the boundary FI term is the 5D
hypermultiplet mass which would lead to the quasi-localization of hypermultiplet
zero modes. Using the expression (\ref{eq:hmass}) for the hypermultiplet mass
in SUGRA and the above result of $\langle\beta\rangle$, one easily finds
\begin{eqnarray}
m_{\Phi_\pm}^2 &=&
\Big( c_w^2 \pm c_w - {\textstyle \frac{15}{4}} \Big) k^2 \epsilon^2(y)
+(3 \mp 2c_w)k \big( \delta(y)-\delta(y-\pi R) \big)\,,
\nonumber \\
m_{\zeta} &=& c_wk \epsilon(y)\,,
\label{eq:hmass}
\end{eqnarray}
where
$c_wk \equiv -\Big( c+q \xi_{FI}/2 \Big)$.

\section{Conclusion}

In this paper, we presented a locally supersymmetric formulation for the boundary
FI terms in 5D $U(1)$ gauge theory on $S^1/Z_2$. We introduced a four-form multiplet
to generate the $Z_2$-odd FI coefficient  $\xi_{FI}\epsilon(y)$ within the 5D
off-shell SUGRA on orbifold. The same four-form multiplier can be used to introduce
the correct bulk and brane cosmological constants for the Randall-Sundrum warped
geometry as well as the hypermultiplet kink masses for quasi-localized matter zero
modes.

We then examined the deformation of vacuum configuration triggered by the FI
terms for the flat ($k=0$) and AdS$_5$ ($k \ne 0$) geometry.
We find that irrespective of the value
of  $k$, the only vacuum deformation is the VEV of
the $U(1)$ gauge scalar field which would yield nonzero kink mass for the charged hypermultiplet.
This is consistent with the result of \cite{Correia:2004pz}
using the superfield formulation of 5D conformal
supergravity.

\subsection*{Acknowledgement}
We thank Tatsuo Kobayashi, Keisuke Ohashi and Stefan~Groot~Nibbelink
for useful discussions. This work is supported by KRF PBRG 2002-070-C00022.


\begin{thebibliography}{99}

\bibitem{Arkani-Hamed:1998rs}
N.~Arkani-Hamed, S.~Dimopoulos and G.~R.~Dvali,
Phys.\ Lett.\ B {\bf 429}, 263 (1998)
[hep-ph/9803315];
%
I.~Antoniadis, N.~Arkani-Hamed, S.~Dimopoulos and G.~R.~Dvali,
Phys.\ Lett.\ B {\bf 436}, 257 (1998)
[hep-ph/9804398];
%
N.~Arkani-Hamed, S.~Dimopoulos and G.~R.~Dvali,
Phys.\ Rev.\ D {\bf 59}, 086004 (1999)
[hep-ph/9807344].

\bibitem{Randall:1999ee}
L.~Randall and R.~Sundrum,
Phys.\ Rev.\ Lett.\  {\bf 83}, 3370 (1999)
[hep-ph/9905221].

\bibitem{Arkani-Hamed:1999dc}
N.~Arkani-Hamed and M.~Schmaltz,
Phys.\ Rev.\ D {\bf 61}, 033005 (2000)
[hep-ph/9903417];
%
E.~A.~Mirabelli and M.~Schmaltz,
Phys.\ Rev.\ D {\bf 61}, 113011 (2000)
[hep-ph/9912265];
%
D.~E.~Kaplan and T.~M.~P.~Tait,
JHEP {\bf 0111}, 051 (2001)
[hep-ph/0110126];
%
M.~Kakizaki and M.~Yamaguchi,
hep-ph/0110266;
%
N.~Haba and N.~Maru,
Phys.\ Rev.\ D {\bf 66}, 055005 (2002)
[hep-ph/0204069];
%
A.~Hebecker and J.~March-Russell,
Phys.\ Lett.\ B {\bf 541}, 338 (2002)
[hep-ph/0205143];
%
K.~Choi, D.~Y.~Kim, I.~W.~Kim and T.~Kobayashi,
hep-ph/0305024;
%
K.~Choi, I.~W.~Kim and W.~Y.~Song,
Nucl.\ Phys.\ B {\bf 687}, 101 (2004)
[hep-ph/0307365].

\bibitem{Kawamura:2000ev}
Y.~Kawamura,
Prog.\ Theor.\ Phys.\  {\bf 105}, 999 (2001)
[hep-ph/0012125];
%
G.~Altarelli and F.~Feruglio,
Phys.\ Lett.\ B {\bf 511}, 257 (2001)
[hep-ph/0102301];
%
L.~J.~Hall and Y.~Nomura,
Phys.\ Rev.\ D {\bf 64}, 055003 (2001)
[hep-ph/0103125];
%
A.~Hebecker and J.~March-Russell,
Nucl.\ Phys.\ B {\bf 613}, 3 (2001)
[hep-ph/0106166];
%
A.~Hebecker and J.~March-Russell,
Nucl.\ Phys.\ B {\bf 625}, 128 (2002)
[hep-ph/0107039];
%
L.~J.~Hall, H.~Murayama and Y.~Nomura,
Nucl.\ Phys.\ B {\bf 645}, 85 (2002)
[hep-th/0107245];
%
T.~Asaka, W.~Buchmuller and L.~Covi,
Phys.\ Lett.\ B {\bf 523}, 199 (2001)
[hep-ph/0108021];
%
L.~J.~Hall, Y.~Nomura, T.~Okui and D.~R.~Smith,
Phys.\ Rev.\ D {\bf 65}, 035008 (2002)
[hep-ph/0108071];
%
R.~Dermisek and A.~Mafi,
Phys.\ Rev.\ D {\bf 65}, 055002 (2002)
[hep-ph/0108139];
%
L.~J.~Hall and Y.~Nomura,
Phys.\ Rev.\ D {\bf 65}, 125012 (2002)
[hep-ph/0111068];
%
H.~D.~Kim, J.~E.~Kim and H.~M.~Lee,
JHEP {\bf 0206}, 048 (2002)
[hep-th/0204132];
%
L.~J.~Hall and Y.~Nomura,
Phys.\ Rev.\ D {\bf 66}, 075004 (2002)
[hep-ph/0205067];
%
H.~D.~Kim and S.~Raby,
JHEP {\bf 0301}, 056 (2003)
[hep-ph/0212348];
%
H.~D.~Kim and S.~Raby,
JHEP {\bf 0307}, 014 (2003)
[hep-ph/0304104].

\bibitem{Scherk:1978ta}
J.~Scherk and J.~H.~Schwarz,
Phys.\ Lett.\ B {\bf 82}, 60 (1979).

\bibitem{Antoniadis:1990ew}
I.~Antoniadis,
Phys.\ Lett.\ B {\bf 246}, 377 (1990).

\bibitem{Randall:1998uk}
L.~Randall and R.~Sundrum,
Nucl.\ Phys.\ B {\bf 557}, 79 (1999)
[hep-th/9810155].

\bibitem{Randall:1999vf}
L.~Randall and R.~Sundrum,
Phys.\ Rev.\ Lett.\  {\bf 83}, 4690 (1999)
[hep-th/9906064].

\bibitem{Choi:2002wx}
K.~Choi, H.~D.~Kim and I.~W.~Kim,
JHEP {\bf 0211}, 033 (2002)
[hep-ph/0202257].

\bibitem{Fujita:2001bd}
T.~Fujita, T.~Kugo and K.~Ohashi,
Prog.\ Theor.\ Phys.\  {\bf 106}, 671 (2001)
[hep-th/0106051];
%
T.~Kugo and K.~Ohashi,
Prog.\ Theor.\ Phys.\  {\bf 108}, 203 (2002)
[hep-th/0203276].

\bibitem{Lin:2003ju}
Y.~Lin,
JHEP {\bf 0401}, 041 (2004)
[hep-th/0312078].

\bibitem{Gherghetta:2000qt}
T.~Gherghetta and A.~Pomarol,
Nucl.\ Phys.\ B {\bf 586}, 141 (2000)
[hep-ph/0003129].

\bibitem{Falkowski:2000er}
A.~Falkowski, Z.~Lalak and S.~Pokorski,
Phys.\ Lett.\ B {\bf 491}, 172 (2000)
[hep-th/0004093].

\bibitem{Ghilencea:2001bw}
D.~M.~Ghilencea, S.~Groot Nibbelink and H.~P.~Nilles,
Nucl.\ Phys.\ B {\bf 619}, 385 (2001)
[hep-th/0108184].

\bibitem{Barbieri:2002ic}
R.~Barbieri, R.~Contino, P.~Creminelli, R.~Rattazzi and C.~A.~Scrucca,
Phys.\ Rev.\ D {\bf 66}, 024025 (2002)
[hep-th/0203039].

\bibitem{GrootNibbelink:2002wv}
S.~Groot Nibbelink, H.~P.~Nilles and M.~Olechowski,
Phys.\ Lett.\ B {\bf 536}, 270 (2002)
[hep-th/0203055];
%
S.~Groot Nibbelink, H.~P.~Nilles and M.~Olechowski,
Nucl.\ Phys.\ B {\bf 640}, 171 (2002)
[hep-th/0205012].

\bibitem{Barbieri:1982ac}
R.~Barbieri, S.~Ferrara, D.~V.~Nanopoulos and K.~S.~Stelle,
Phys.\ Lett.\ B {\bf 113}, 219 (1982);
%
A.~H.~Chamseddine and H.~K.~Dreiner,
Nucl.\ Phys.\ B {\bf 458}, 65 (1996)
[hep-ph/9504337];
%
D.~J.~Castano, D.~Z.~Freedman and C.~Manuel,
Nucl.\ Phys.\ B {\bf 461}, 50 (1996)
[hep-ph/9507397];
%
P.~Binetruy, G.~Dvali, R.~Kallosh and A.~Van Proeyen,
hep-th/0402046.

\bibitem{Green:sg}
M.~B.~Green and J.~H.~Schwarz,
Phys.\ Lett.\ B {\bf 149}, 117 (1984).

\bibitem{Marti:2002ar}
D.~Marti and A.~Pomarol,
Phys.\ Rev.\ D {\bf 66}, 125005 (2002)
[hep-ph/0205034].

\bibitem{Abe:2002ps}
H.~Abe, T.~Higaki and T.~Kobayashi,
Prog.\ Theor.\ Phys.\  {\bf 109}, 809 (2003)
[hep-th/0210025].

\bibitem{Bergshoeff:2000zn}
E.~Bergshoeff, R.~Kallosh and A.~Van Proeyen,
JHEP {\bf 0010}, 033 (2000)
[hep-th/0007044].

\bibitem{Correia:2004pz}
F.~P.~Correia, M.~G.~Schmidt and Z.~Tavartkiladze,
hep-th/0410281.

\bibitem{PaccettiCorreia:2004ri}
F.~Paccetti Correia, M.~G.~Schmidt and Z.~Tavartkiladze,
hep-th/0408138.

\bibitem{Abe:2004ar}
H.~Abe and Y.~Sakamura,
JHEP {\bf 0410}, 013 (2004)
[hep-th/0408224].

\bibitem{nonintegrable:FI}
H.~Abe and K.~Choi, in preparation.

\bibitem{Ceresole:2001wi}
A.~Ceresole, G.~Dall'Agata, R.~Kallosh and A.~Van Proeyen,
Phys.\ Rev.\ D {\bf 64}, 104006 (2001)
[hep-th/0104056].




\end{thebibliography}
\end{document}